\documentclass[sigconf,anonymous=false]{acmart}

\usepackage[draft]{todonotes}

\AtBeginDocument{%
 \providecommand\BibTeX{{%
  \normalfont B\kern-0.5em{\scshape i\kern-0.25em b}\kern-0.8em\TeX}}}

\setcopyright{acmcopyright}
\copyrightyear{2022}
\acmYear{2022}
\acmDOI{}

\acmConference[NSPW '22]{NSPW '22: New Security Paradigms Workshop }{October 24--27, 2022}{New Hampshire, USA}
\acmBooktitle{NSPW '22: New Security Paradigms Workshop ,
 October 24--27, 2022, New Hampshire, USA}
\acmPrice{}
\acmISBN{}

\begin{document}

\title{Transparency, Compliance, And Contestability When Code Is(n't) Law}


\author{Alexander Hicks}
\email{alexander.hicks@ucl.ac.uk}
\affiliation{%
 \institution{University College London}
 \streetaddress{}
 \city{}
 \state{}
 \country{}
 \postcode{}
}

\renewcommand{\shortauthors}{Alexander Hicks}

\begin{abstract}

Both technical security mechanisms and legal processes serve as mechanisms to deal with misbehaviour according to a set of norms.
While they share general similarities, there are also clear differences in how they are defined, act, and the effect they have on subjects.
This paper considers the similarities and differences between both types of mechanisms as ways of dealing with misbehaviour, and where they interact with each other.

Taking into consideration the idea of code as law, we discuss accountability mechanisms for code, and how they must relate to both security principles and legal principles.
In particular, we identify the ability to contest norms enforced by code as an important part of accountability in this context.
Based on this analysis, we make the case for transparency enhancing technologies as security mechanisms that can support legal processes, in contrast to other types of accountability mechanisms for code.
We illustrate this through two examples based on recent court cases that involved Post Office in the United Kingdom and Uber in the Netherlands, and discuss some practical considerations.
\end{abstract}




\maketitle

\section{Introduction}
Computer systems now have a broad, and increasing, role in people's lives, even when they do not interact with or have any privilege over these systems.
The code that makes up these systems defines what these systems do and, therefore, the norms that they apply when functioning.
The impacts of applying these norms can be negative and unfair, because they result from systems that are flawed e.g., unreliable or discriminatory, or that, by optimising certain performance metrics at the cost of fairness, welfare, and other values, produce harmful externalities.

The harms of such system are reserved for those that are subject to the system, and may not affect the entities that design and operate these systems -- they may even benefit in some cases.
In other words, the costs associated with these harms are externalised to the public, which has little recourse to mitigate these harms.
This makes dealing with this source of harms difficult.

Typically, harms done to people are addressed through the law and the legal systems that enforce it.
This allows victims of harm to be compensated, and misbehaviour that results in harms to be punished and disicentivised.

Security, on the other hand, typically deals with ensuring that systems function as intended i.e., that they are reliable and unlikely to be compromised by an adversary.

When dealing with harms that result from the application of code, both of these fields come into play.
Law should ensure that victims of code enabled harm should be able to contest the systems that cause these harms.
Security should ensure that people should not fall victim to flawed systems.

In practice, however, this currently does not work.
There are widespread issues with people suffering from flawed systems that have been applied to determine, among other applications, entry to buildings via facial recognition, jail sentences, and so on. 
Systems evolve quickly and the law has not kept up with the application of technology to these aspects of our lives, allowing harm to occur without sanction, and making it hard for victims of harm to contest the application of code enforced norms that have caused harm.

To deal with this issue, the idea of algorithmic accountability, which studies how to design ways of making algorithms accountable, has gained in popularity.
In line with this field of work, this paper works towards addressing the issue of reconciling the use of security mechanisms that can assert the behaviour of a system with legal processes that can be used to contest the norms enforced by a system.

We begin with the idea of norms, misbehaviour, how legal processes and security mechanisms work as two ways of dealing with misbehaviour, and the interaction between these two approaches.
Looking at this through the lens of code as law and digisprudence, we argue both the need for secure accountability mechanisms and how they must be designed to make them useful as tools to contest code enforced norms.
This allows us to compare different approaches to accountability, and make the case for transparency enhancing technologies against less transparent forms of audits based on assurances of compliance with norms.
We illustrate this through two examples based on recent court cases that involved Post Office in the United Kingdom and Uber in the Netherlands.
We also discuss some practical considerations that relate to electronic evidence, balancing transparency and privacy, and where transparency should be implemented with respect to the system it acts on.

\section{Preventing Misbehaviour Through Legal Processes and Security Mechanisms}

\subsection{Norms and misbehaviour}\label{sec:norms}
Misbehaviour is the action of deviating from a norm.
Following Hildebrandt's discussion of legal and technological normativity~\cite{hildebrandt2008legal}, we think of norms as regulative (mandating, permitting, or disallowing some pre-existing possible action) or constitutive (defining a possible action).
The difference between both can be thought of in terms of how misbehaviour can occur in each case.

In the case of regulative norms, misbehaviour can occur by deviating from the regulative norm at hand by performing a disallowed action, which does not prevent the action from being performed but does entail possible punishment.
For example, a regulative norm may stipulate that car should not be driven over a specified speed limit.
This does not prevent driving a car at a higher speed (this is the driver's choice to make) but can lead to penalties enacted by the relevant authority if the speed limit is exceeded.

In the case of a constitutive norm, deviating entails not performing an action defined by the norm and, therefore, the expected result of adhering to the constitutive norm at hand does not occur, resulting in a form of failure for the user and the state of the system remaining unchanged.
For example, a constitutive norm may be the rule that a password must be entered to login into an account.
If no password, or the wrong password, is entered, then it is simply not possible to login into the account - the user has no choice but to enter the correct password or they will fail to login and the state of the system will not change as the user's status will not change if they stay logged out.

There is of course the question of who defines what the norms are, and thus misbehaviour. 
For computer systems, the system's code often defines constitutive norms as it creates actions related to the system that did not exist before the system.
(Different systems may of course share similar mechanisms and even re-use code e.g., a login mechanism, but logging into one system is not the same action as logging into another system.)
Thus, whoever designs and implements (and in the case of a data driven system, trains) the system has significant power to determine the norms that are put in place by the system.


The code itself is also the result of the social norms and practices of those who write it.
This results in an expected behaviour model (explicitly specified or not) that the user should follow, with anything that deviates in a relevant way from this expected behaviour being thought of as misbehaviour.
For example, in the US, the flawed design and training of an algorithm that produces risk assessments used to help determine whether an imprisoned person should be released resulted in black defendants being incorrectly labelled as higher risk compared to white defendants who were incorrectly labelled as lower risk~\cite{angwin2016machine}.
This determined that, even if this was not intended, being black constituted a deviation from the expected defendant model and was punished with harsher sentences -- a reflection of a system that already disproportionately imprisons black Americans~\cite{nellis2016color}.


As we have seen, norms that are put in place by a system, as well as those that form the context in which the system was created and is operated, influence how that system functions and the experience of its users.
They also influence the way the system is designed to respond in cases where it determines that users have deviated or misbehaved in some way.
Our interest is in the design of mechanisms for the mitigation of harms that could result from the use of a system.
This too will need to be informed by an understanding of how norms are put in place in technology and how they can be challenged (and changed).

\subsection{Law based disicentivisation and punishment of misbehaviour}
Law broadly defines the limits of acceptable behaviour and the consequences of unacceptable behaviour in everyday life. 
Its purpose is twofold.
First, it disincentivises people from acting in a manner that is defined as unacceptable by the law.
Second, if people nonetheless act in such a manner, the law makes it possible to punish them via legal processes that are themselves defined in law.
These processes rely on the availability of admissible evidence that shows beyond a certainty threshold that the person to be punished did in fact act unacceptably.

Processes are more fundamental to the law than specific laws are themselves.
(Of course, processes are usually defined in the law itself, but we distinguish here between laws that are applied to determine the resolution of the question that results in a legal process from the laws determining how the legal process proceeds.)
Although both vary across jurisdictions and are mutable, new laws that determine acceptable behaviour are introduced and changed much more frequently than the processes that are used to adjudicate e.g., trials.
Moreover, the regular changes in laws show that they can be contested, as do the interpretation of the laws themselves, which is determined by courts and may be the subject of legal processes themselves.

The state institutions that legislate, enforce, and adjudicate laws are typically well defined, although they vary across states.
It is also possible for other organisations, such as private businesses, to act as rule makers and enforcers over the jurisdiction of a system they operate, for example through terms of service agreements, although these may, in turn, be subject to state enacted regulations and the states legal system that would handle any dispute.

\subsection{Security against threats and a posteriori security}
Security, or more precisely information security in our context, works by defining mechanisms based on a defined threat model. 
There is no notion of absolute security, only security against a given threat model that relies on specified assumptions about the capabilities of an adversary and the difficulty (in a computational sense) or cost (in an economic sense) of performing certain tasks.

Traditionally, security mechanisms are constitutive and impose behaviour that honestly follows a protocol, implying that an adversary cannot possibly misbehave and break the security guarantees provided by the security mechanism (otherwise the mechanism would not be secure by definition).
For example, a provably secure encryption scheme that is well implemented cannot be broken by an adversary with more than a negligible probability (in the formal mathematical sense of the term).
Thus, security mechanisms are very different from how regulative legal mechanisms function.
Misbehaviour cannot happen in principle and, therefore, there is no kind of accountability process or defined punishment for the adversary.

Not all misbehaviour by individuals or algorithmic systems can be stopped a priori, however, which has motivated work on security mechanisms designed instead to detect misbehaviour and produce evidence of that misbehaviour~\cite{weitzner2008information}.
Moreover, the reliability of preventative security mechanisms must also be empirically examined on occasion.
In the context of this paper, this type of a posteriori security mechanisms is what we focus on.

An example of this being successfully deployed in practice is Certificate Transparency~\cite{laurie2014certificate}.
This is a now widely adopted~\cite{8835212} system that provides tamper-evident transparency logs that record the issuance of SSL certificates for websites by certificate authorities.
Certificates are either logged, making them easy to inspect, or not logged in which case a browser client that encounters such a certificate can report it.
This allows misbehaving certificate authorities who (on purpose or due to compromise~\cite{van2013diginotar}) emit problematic SSL certificates to be detected, disincentivisng such misbehaviour or, conversely, incentivising security on the part of certificate authorities.

The consequences of misbehaviour that is easily detected can be severe if sanctions are imposed.
A certificate authority that is deemed to have misbehaved by Google, Mozilla, Microsoft, and other browser vendors may be blacklisted by their browsers, mitigating the harm done to users, and practically ensuring that the certificate authority quickly goes out of business.
For example, DigiNotar went bankrupt shortly after being compromised and having its certificates deemed untrustworthy~\cite{digibankrupt}.

Security mechanisms like Certificate Transparency are defined as technical mechanisms that record evidence of misbehaviour and, therefore, function more like regulative mechanisms, there is no built-in notion of accountability process or punishment.
That is left to whoever relies on these mechanisms, such as Google (who dominates the browser space~\cite{browsershare}) and other browser vendors.

Thus, unlike law, it is the technical mechanism that is fundamental here, rather than the process of dealing with misbehaviour once it is detected.
More often than not, there is no well defined accountability process and it is instead determined by power relations around the system.

This implies a distinction between the security of the system (e.g., making sure that certificates are trustworthy) and the security of the parties in the system.
DigiNotar vanished following its security incident, browser vendors protected their product and users from future harms, but those affected by illegitimate certificates before measures were taken were not protected or compensated in any way due to this mechanism.

\subsection{Economic considerations}

Economics considerations play a role in both cases.

Harms caused by algorithmic systems often do not fall under criminal law and, therefore, the consequences are primarily financial, which leads to economic considerations of expected costs.
As Wu puts it ``laws impose costs upon regulated groups. Those groups that seek to minimize the costs of law face a fundamental choice between mechanisms of change and avoidance. Both mechanisms have the effect of lowering the expected costs of law, but the similarities end there. Mechanisms of change (principally lobbying) decrease the sanction attached to certain conduct and tend to require collective action. Mechanisms of avoidance, on the other hand, decrease the probability of detection and typically do not require that groups act collectively, but depend on specific vulnerabilities in the law.''~\cite{wu2003code}.
For example, Google has multiple times paid fines to the European Union that are greater than many of the contributions to the European Union by member states~\cite{google}.

Similarly, the security of systems often relates to the underlying economics of securing the system~\cite{anderson2001information}.
Securing a system has a cost that, economically speaking, is only worth expending if it outweighs the expected loss due to successful exploitation of the system by an attacker or, more generally, a system fault.
When the costs of system fault can be passed on to the users who are harmed, there is a perverse incentive not to expend resources on making the system reliable.

The economic considerations related to both legal mechanisms and security mechanisms directly relate to each other when we consider cases where a system can harm users.
If that is the case, the legal risk for the system operator is that they may be legally responsible for the harm done to users by faults in the system.
The European GDPR, for example, makes these regulatory risks real for certain kinds of data protection failures.
In these cases, the economic considerations of expected costs due to the risk of regulatory non-compliance are the economic considerations that can favour (or not) the implementation of reliable security mechanisms.

\subsection{The interaction between security mechanisms and legal mechanisms}~\label{sec:interaction}

The overlap between legal processes and security mechanisms happens when a security mechanism is intended to ensure compliance with a legal norm.
A common historical example of this taking place is the repeated attempts to apply copyright and digital rights management (DRM) to online content, which motivated both technical work (see, for example, Chapter 24 of Anderson's book~\cite{anderson2020security}) and legal work on the interaction between code and law~\cite{lessig2003law}.

This put the focus on two things.
First, technology could change the efficacy of a law and facilitate unwanted behaviour e.g., distributed online file sharing made it much easier to ignore intellectual property law.
Second, technology could be used to deal with the change in the efficacy of a law by deploying mechanisms that are secure against the unwanted behaviour enabled by technology e.g., DRM mechanisms.

While security mechanisms and legal mechanisms are both ways of enforcing norms and interact in many situations, they are not interchangeable.
Security mechanisms, in particular a posteriori security mechanisms, are technical mechanisms that enable the collection of evidence. 
Legal mechanisms are processes of determining the consequences that should be applied to parties in response to their behaviour based on evidence related to that behaviour.

Thus, the legal analogy for a posteriori security mechanisms is that of evidence collection while the accountability process is in the hands of those who can (i) access that evidence (which may be determined by technical access control mechanisms) and act upon it (which requires agency and authority).
Re-iterating on the previous example of Certificate Transparency, while everyone can monitor Certificate Transparency logs, it is effectively only browser vendors who can act upon the information they contain and enact some kind of accountability on the misbehaving certificate authority.
Although this may manifest itself through code e.g., blacklisting certificates signed by the misbehaving certificate authority, the process of accountability is a decision process within the organisations themselves, not one determined independently by code.

The interaction between security mechanisms and legal mechanisms for accountability is, therefore, centred on how security mechanisms can be leveraged to serve legal mechanisms.

This interaction is not necessarily frictionless, however, as there can be a ``clash between rules and principles exacerbates the difference in perspective between system designers, who favour formal rules, and policy makers, who are more comfortable with situational application of principles''~\cite{feigenbaum2018incommensurability}.
Unlike Google with Certificate Transparency, the legal system and many more organisations do not have both the capacity to design and make use of technical mechanisms that can support accountability processes.
Without such capacity, however, dealing with systems that can produce harms is difficult.


\section{Accountability Through The Lens of Code Is Law and Digisprudence}

\subsection{Code is Law and Digisprudence}

The notion of code as law in academic work goes back to Reidenberg~\cite{reidenberg1997lex} who noted that ``technological capabilities and system design choices impose rules
on participants'' and Lessig ~\cite{lessig2000code} who framed the issue as ``we therefore don't see the threat to liberty that this regulation presents''.

The use of code as a part of legal actions existed as transactions tied to contracts were already being executed through code at that time.
Moreover, Szabo had introduced the idea of smart contracts~\cite{szabo1997formalizing} that made explicit the possibility of contractual transactions that would execute entirely through ``smart contracts'' implemented in code.
Smart contracts are now the basis for cryptocurrencies such as Ethereum that are essentially decentralized smart contract platforms~\cite{wood2014ethereum}, and law scholars have studied their role as legitimate legal contracts~\cite{raskin2016law}.

More generally, however, code that defines the operation of technical systems forms, like law, a way to regulate the behaviour of people subject to the system.
Subjects of the system in this case include not only people operating the software or are users of the system, but also those on whom the system can have an effect.
For example, someone who is run over by an autonomous vehicle operating software that did not determine the vehicle should stop once the person was identified will be affected by the software operating the car without ever interacting with it or consenting to be subject to it.
This effect can also be mediated by a third party, including in legal matters, as is the case when judges make decisions based on the outputs of (generally biased) automated decision making systems~\cite{angwin2016machine}.

As Diver~\cite{diver2021digisprudence} suggests, code is not law per se, even if its automation means that it governs the behaviour of people in the system in a more effective way, because it lacks law's mechanisms of ex-ante legitimation and ex-post remediation.
Diver makes four claims about code, its effect, and its design~\cite{diver2021digisprudence}.

First, code can have regulative effects on behaviour that are more pervasive and direct than law is capable of.
Moreover, the regulatory effects of code do not need to be compatible with law.

Second, norms that regulate citizens, including those that are imposed by code, ought to be legitimate in that they ensure certain formal qualities in their design.

Third, attention should be paid not only to the legitimacy of code but also to the legitimacy of the design of code.

Fourth, legitimation of a code imposed rule must be done at design time because there is little scope to re-interpret code after the fact.

Dealing with this requires a theory of what constitutes legitimate code that Diver names digisprudence, which is based on the following affordances: transparency about the provenance, purpose, and operation of code; oversight; choice; intelligibility supported by delay; and contestability as the overarching concern~\cite{diver2021digisprudence}.

\subsection{Digisprudence and Accountability}

Digisprudence as a framework is aligned with the desire for accountability mechanisms that can provide the affordances we have just listed, beginning with transparency about the provenance, purpose, and operation of code.
Oversight is required to make use of transparency in order to apply accountability.
Choice is related to the norms enforced by the system, or simply the choice to be subject or not to these norms, which requires transparency about these norms and how they are applied in the first place.
Intelligibility and the affordance of delay are, in turn, required for oversight and choice to take place.

Contestability is less integral to the discourse about accountability.
For example, Wieringa's recent systematic review of the field does not mention contestability~\cite{wieringa2020account}.
Rather, accountability is often focused on whether or not a system has functioned correctly instead of the legitimacy of the norms the system applies -- ``trust but verify'' as the saying goes (see Desai and Kroll for example~\cite{desai2017trust}).
This suggests that a choice must be made between wanting accountability for the performance of the system (which does not require contestability) or accountability for the norms enforced by the system (which requires contestability).
We return to this in the next section.

Because the accountability mechanisms we are concerned with here also involve code and, indeed, accountability mechanisms are designed to apply norms, we must also pay attention to how these affordances are taken into account when designing and executing accountability mechanisms.

Fundamentally, accountability mechanisms must reveal information about the system and enable action to be taken based on that information (which may include legal action or some other process).
Thus, they regulate access to information and the effects of access to that information.

The provenance, purpose, and operation of an accountability mechanism should make clear what the mechanism is intended to provide accountability for, to whom, and how.
The incentives of the party that designs the accountability mechanism are important.
An accountability mechanism designed for a system by the system's operator that primarily works to prove the correct execution of the system may, for example in zero-knowledge\footnote{A zero knowledge proof is a cryptographic proof of a computational statement that reveals nothing but the proof of the statement e.g., proving the correct execution of a computation.} as suggested by Kroll et al.~\cite{kroll165accountable}, not be considered as legitimate by the public as another mechanism for the same system that reveals more information about not only the system it provides transparency for but also itself.
For example, a zero-knowledge proof, even if publicly verifiable, that is verified by a judge not allow for any explanation beyond ``computer says yes'' or ``computer says no'', which may not be a satisfying explanation for the behaviour of complex systems.

More generally, the assumptions that underpin the design play an important role because they can determine the legal effect of the use of the accountability mechanism (e.g., it supports the production of admissible evidence to be used in court) but also the type of misbehaviour that it can provide accountability for based on the threat model (e.g. whether the system operator and code are considered adversarial to accountability) that determines its security design.

Assumptions about the code that is subject to accountability are also important.
Interpreting code as law generally entails considering code as a form of strong legalism, but this assumes that the code is reliable and secure, otherwise its effects can be bypassed and it fails to demonstrate strong legalism.
Accountability mechanisms must take this into account by not assuming that the code is necessarily reliable and secure, and by being designed to function independently of the code so that it does not fail if the code fails.

There should be oversight over the use of accountability mechanisms, to make sure that they are effective in providing accountability, and that the way they regulate access to information and the effects of access to that information is aligned with its design and purpose.
Of course, intelligibility (or usability in the context of designing a secure accountability mechanism) is necessary for this to be possible.

Likewise, choice must be given to be subject to the norms accountability mechanisms entail.
Either for the system operator whose system will be subject to an accountability mechanism, in the case where there are no regulations requiring its use. (If there are regulations, there is a notional choice to abide by them and flexibility in how to implement them.)
This point has been made under the guise of protecting commercially sensitive aspects of the system~\cite{desai2017trust,kroll165accountable}.
This also applies to users of the system whose information may be revealed as part of transparency.

Contestability also matters, because accountability mechanisms should enable consequences.
The fact that, for example, transparency by itself is not always effective is that it can fail to enable further actions~\cite{ananny2018seeing,doi:10.1111/j.1467-9760.2010.00366.x}.
Thus, it should be possible to contest accountability mechanisms so that the consequences (or lack of consequences) can be considered legitimate.
A practical example of this is for mechanisms that serve as evidence producing mechanisms that enable legal dispute, the admissibility of the evidence produced can be contested according to the norms set out of law that regulates evidence.
We explore this in greater detail in the next sections.

\section{From Accountability to Contestability}





In Section~\ref{sec:interaction} we highlighted a takeaway from the interactions between security mechanisms and the law, which is that technology can (i) serve to bypass and (ii) enforce law.
If we take code as acting somewhat like law, this is still true. 

Hacking, Privacy Enhancing Technologies (PETs), and Protective Optimisation Technologies (POTs)~\cite{kulynych2020pots,gurses2018stirring} show the existence of this interaction in practice.

Hacking attempts to do something that is not allowed by the norms of the system.
This is often viewed through the lens of criminal hacking, but it can also fall in grey legal areas~\cite{evtimov2019tricking} or be done to contest norms that are reasonably considered illegitimate. 
In general, this is a solution that does not scale well because it can require technical skills that are not widespread among users, and typically does not entail any modification of the hacked system that would benefit users other than the hacker.
An example where this is useful, however, is when it prevents the system from functioning (if this is not outweighed by some benefits the system might bring) or leads to greater transparency about the system (like whistleblowing) that can be leveraged to contest the system.

PETs constrain the capability of code that is designed to collect private information.
For example, end-to-end encryption, which is widely deployed in messaging services, prevents the ability for someone to execute code that would eavesdrop on a conversation, which would otherwise be possible by default.
After more than twenty years since PETS became an active topic~\cite{goldberg1997privacy}, privacy engineering is now its own discipline~\cite{gurses2011engineering,gurses2015engineering} backed by data protection regulations (e.g., the GDPR), although systems still routinely compromise user privacy to satisfy a logic of information accumulation and surveillance~\cite{zuboff2015big}.

POTS attempt to overrule the effects of code driven optimisation, allowing users outside of the system to intervene without requiring cooperation from the system's operator.
For example, using Sybil devices to generate fake traffic in an area can stop traffic routing apps (e.g., Waze) from routing traffic to the area and mitigate the negative externalities that would otherwise ensue in said area~\cite{kulynych2020pots}. 
There is, however, no guarantee that such interventions cannot in turn be optimised away by the target system once it is adjusted to take the existence of a POT into account.

These tools are available to individuals and can be effective (even if only to a limited extent) against code designed and deployed by states, companies, and other large institutions, showing that contesting code imposed norms is sometimes possible (although these tools are not necessarily accountability mechanisms).
Code not only enforces norms but it can also be used to contest and bypass norms, and the fact that these tools are user centric distinguishes contestability from traditional accountability that is centred on the system operator.

When code imposed norms are discussed and determined to be harmful in some sense, through the use of secure accountability mechanisms, they can be changed.
Even in the case of code that is intended to provide immutability by design, such as blockchains, these guarantees are void if other interests are deemed more important.
Following the loss of $36,000,000$ ether due to an insecure smart contract, Ethereum users simply decided to fork Ethereum to revert the situation~\cite{kiffer2017stick}, creating Ethereum Classic (which did not revert the hack) and Ethereum (which did).
Ethereum, the forked chain that decided that ``code is law'' was not worth it at that moment, has since been the dominant chain .

How did this happen?
The realisation that the loss of funds was (i) of great value, both financially, and in terms of the ability for users to trust the system with their funds; (ii) reversible because it was possible to introduce code that would transfer the stolen funds back to their original owners, at the cost of forking the chain; (iii) reverting the hack was supported by many powerful members of the community e.g., Vitalik Buterin (Ethereum's most important public figure~\cite{ripvitalik} and idea contributor~\cite{azouvi2018egalitarian}).
Ethereum was, therefore, clearly accountable to its users, who in turn were able to contest -- at least those that had more influence over the community -- the norm applied to their ability to recover funds.
Moreover, because of the transparency offered by Ethereum, it was possible for any interested user to see exactly what had happened, what could be done, and what was done in the end.

This example shows that transparency enabled accountability can be used to contest the effects of code and change them.
This result is not necessarily generalisable, however, because it played out in favour of those with disproportionate power over the system.
In many cases where we would like to introduce accountability to the extent that norms can be contested, those with power over the system (e.g., system operators) are not those that wish for the norms to be contested.
Rather, they are those who want to enforce these norms in the first place.
This brings back a common theme with accountability, which is the importance of power relations around systems.

This should inform how we design accountability mechanisms because, as mentioned in the previous section, accountability mechanisms can regulate the effects of access to the information that the mechanism makes available to some.
This is because the format of that information plays a role in how it can be used.
If any aspect of the system is to be contested, therefore, it must be determined how this will happen.

Some systems, such as Ethereum in the example above, afford more power to their user communities but this requires a level of decentralised governance that is rare.
For the vast majority of systems i.e., systems deployed by centralised private entities, there are no governance mechanisms that could allow an individual subject to the system to systematically influence it.
Thus, in this paper we will focus on contesting norms enforced by systems through legal processes with the intent of contesting the formulation of these norms as, although far from ideal, the legal system is often the best chance of achieving this an individual will have.
As a result, the format of the information that accountability mechanisms provide should be usable as part of public disclosures of information about the system and admissible evidence to be used in court to support an argument in a dispute about the system.

\section{Compliance And Transparency Based Auditing}






\subsection{Verification and compliance based auditing}

In theory, systems could be formally verified and, therefore, treated as reliable assuming that no design flaws were presented (a strong assumption in itself).
In practice, however, formal verification tools are of limited use because many systems involve multiple different protocols that interact with each other across different hardware, software, and network conditions, making formal verification of an entire system unrealistic.

Software is often continuously modified (as well as the operating system it runs on), in particular for new applications, can involve millions of lines of code representing extremely complex protocols, with a non-zero rate of bugs in the code and logic flaws at the design level.
Data is shared across networks operated by different parties, in varying network conditions (affecting reliability or synchrony assumptions required by distributed protocol design models), which makes strict enforcement mechanisms impractical~\cite{weitzner2008information}.
Even hardware, at a scale at which some large scale applications operate, may fail to be reliable for basic tasks such as encryption and decryption~\cite{hochschild2021cores}.

A weaker form of verification that is more realistic is based on compliance based auditing that checks the correct execution of a process in a system rather than the correctness of the system itself.
For example, automated tools may work by checking for compliance with certain norms e.g., certain specific clauses of the GDPR~\cite{arfelt2019monitoring}.
This is limited to cases where the desired norm is assumed, or simply required by law, which may not always be the case.
In practice, many systems enforce norms that fall under a grey legal area, or like many clauses of the GDPR, are not related to compliance, system behaviour, or require interpretation, and cannot be encoded in logic and automatically checked for compliance.

The automated aspects of these tools do not provide any agency to any individual that would be harmed by the system, because there is no need for them to provide access to any information to unprivileged users of the system.
For example, a system operator may be able to show that the system has complied with the desired norms when it has, but when it hasn't a user may not be able to generate any evidence of this.
This solution, therefore, benefits honest system operators but does not necessarily punish those that operate flawed systems.

Moreover, because the focus is on compliance with a pre-established norm, it does not leave much space to discuss the norm itself.
A logical compliance test that returns a boolean pass/fail value will not be able to provide much information about edge cases or the cause of passes or fails that may be necessary to evaluate the norms, and the reason that the system satisfies a norm may be that the norm itself is specified erroneously.
Having a human in the loop also brings its own challenges~\cite{green2021flaws}, and may risk the humans in the loop legitimising a system because it passes compliance checks that do not represent all the harms they may cause.

It is also important that the software be well designed to represent the norm it wishes to verify and secure enough to operate in an adversarial environment.
For example, Volkswagen developed software that could detect when their cars were being tested so that they could change their performance accordingly~\cite{volks}.

Hardware that supports trusted execution environments and cryptographic tools that can be used to verify computations~\cite{parno2013pinocchio} can be applied to verify the execution of the assurance software can be applied in cases where the threat model requires it and to permit public verifiability.
For example, methods of providing the public with cryptographic proofs that certain processes have followed have been proposed, based on zero-knowledge proofs and secure multiparty computation~\cite{goldwasser2017public,frankle2018practical}.
Although the outputs of these systems can be verified, their inputs cannot.
Thus, this amounts to assuming honesty on the part of those that control the inputs and, therefore, that the processes that are meant to be audit have been followed correctly.
This is not an appropriate threat model for many cases where it can be assumed that processes may not be followed honestly and systems may be faulty.
Moreover, because zero-knowledge proofs obfuscate practically all information, their use is very limited to investigate misbehaviour that would involve nuanced details~\cite{stark2021certificate}.

Finally, as Kim points outs~\cite{kim2017auditing} transparency and audits are still necessary even if assurances exist, because the fault in the system that causes harm may not be in the code but in the design itself.



\subsection{Transparency Enhancing Technologies}

Transparency Enhancing Technologies, in contrast to compliance based solutions, focus on making information about the system available rather than evaluating the system. 
The evaluation is regarded as another process (which may or may not be automated) that is therefore more transparent because the information it is based on is more widely available.

In terms of technical mechanisms, this approach is therefore based on creates logs of operations in the system, for which there are well defined cryptographic security models~\cite{chase2016transparency} as well as implementations of reliable logs (e.g., the Certificate Transparency logs).
Kroll provides a survey of traceability mechanisms~\cite{kroll2021outlining}.
Given a log of a program's actions in the system, it may also be possible to determine the program actions that were actual causes of the program deviating from its specified behaviour~\cite{datta2015program}.
Likewise, for machine learning based systems, it can be possible to quantify the degree of influence of inputs on outputs of the system and release the information for transparency~\cite{datta2016algorithmic}.

Transparency is based on recording and making information available, therefore, it does not assume a norm for the system like compliance based solutions.
Thus, it makes it possible to explore what that norm is via the information it makes available.
Moreover, it does so independent of the system's norm that may have been specified at its design stage.
This is akin to adopting a stronger threat model that makes fewer assumptions about the system it audits and those that interact with the system.
It can, therefore, identify discrepancies between norms that were desired at the design stage and those that are actually enforced as the system operates.

Transparency can also be more public facing and democratic than compliance based solutions.
First, it is based on releasing information rather than checking it.
Second, a transparency system (e.g., logs) can be maintained by various parties and relied on by others.
Assurance software, however, must be possessed by those who execute it and are typically not publicly available.
A broader audience invites a broader critique.

An example comparison between an compliance based system and a transparency focused system can be made in this case.
We have already mentioned the work of Frankle et al.~\cite{frankle2018practical}, which uses cryptographic tools (zero-knowledge proofs and multiparty computation) to verify that secret legal processes to authorise surveillance, for example, have been followed.
The output of this solution is a cryptographic proof that processes have been well followed, and statistics about these processes, but it does not reveal anything else.

Another paper by Hicks et al.~\cite{hicks2018vams} addresses a similar problem, that of auditing requests for access to data made by law enforcement.
This paper proposes a solution that logs (similarly to Certificate Transparency) and releases the log of requests for access to data with read access reserved to auditors (for all requests) and individuals (to see requests for their data).
This allows publicly verifiable statistics to the extent that individuals can verify the inclusion of requests for their data in the computation of the statistics, and recompute the statistics themselves based on a privacy preserving synthetic dataset.

The first system, proposed by Frankle et al. offers stronger confidentiality guarantees, but is only useful if processes are followed correctly.
If they are not, not much can be learned by design.
The second system offers confidentiality guarantees that are weaker than those offered by zero-knowledge proofs because more information is revealed by the release of a synthetic dataset of logged requests.
However, it can be used to identify errors (i.e., deviations from the specified ``honest'' norm) and abuse (i.e., the existence of a malicious norm) more effectively, and with greater agency for those affected.
Thus, if things go wrong, this solution may be more useful in contesting the system it looks at.

There are, therefore, trade-offs to consider, but if the ability to contest norms is required then the argument is in favour of transparency that can accurately produce evidence of the system producing behaviour that does not respect the desired norm, or correctly enforcing a harmful norm.

\subsection{Examples of the usefulness of system transparency in court cases}
\paragraph{Post Office Limited and its unreliable accounting system}
Post Office Ltd is a state owned private company in the United Kingdom (UK) that provides a variety of services to customers including postal and financial services.
Subpostmasters operate Post Office branches on behalf of Post Office Limited and are responsible for any losses at their branch.
The accounting at each branch, however, was handled by a centralised accounting system named Horizon, which was developed in the nineties.
As it happens, Horizon, like most large IT systems, suffered from bugs that could lead to accounting errors.
Over the years, Subpostmasters were accordingly requested to cover the losses or be criminally prosecuted.\footnote{See Nick Wallis' book~\cite{pobook} on the subject for more details.}

One important factor in these prosecutions was the legal presumption in the UK that, unless there is evidence of the contrary, the evidence produced by a computer was reliable.
Post Office had access to a Known Error Log but did not disclose its contents~\cite{mccormack2016post}, and because evidence was treated on a case by case basis, it was never possible to establish the unreliability of Horizon for a single defendant with limited resources.
Thus, ``a subpostmaster could be held responsible for losses they incurred as a direct result of failing to notice an error in a sophisticated computer system over which they had no control''~\cite{mccormack2016post}.

More recently, however, a Group Litigation that allows a collection of cases to be examined in parallel took place.
This allowed subpostmasters to contest Horizon as a group with pooled funds and more combined evidence to contest Horizon more effectively.
As a result, it was possible to force more disclosures about Horizon that made it possible to establish that it was unreliable, with forced the government to put aside hundreds of millions of pounds to cover the payouts in what is considered the biggest single miscarriages of justice in British history~\cite{popayout}.

It is instructive to consider this example, and how similar situations could be improved because it is a large system but one that is nonetheless less complex than, for example, machine learning based systems. 
It also a typical kind of system that people will interact with daily.
Many other faulty traditional systems have caused legal issues~\cite{bellovin2021seeking}.
Reasoning about the responsibility of individual bug occurrences in a system is difficult because if the probability of a bug occurring is similar to the probability of a user committing fraud then we are left with biases~\cite{hicks2019transparency}.

As the Group Litigation showed, an approach based on transparency of the know error log and intelligible recordings of the system's operations could improve things by making accessible the information that was actually useful in practice~\cite{hicks2019transparency}.
This would make it possible for subpostmasters to (i) be aware of potential bugs (transparency about the system), (ii) analyse the logs of their system's operation (transparency about their interaction with the system), and (iii) have access to evidence that can be used to contest any faults that may occur.
Moreover, the security of such a system should be based on a threat model that assumes Post Office to be adversarial to transparency as they actively hid the contents of the known error log.

Relying on (zero-knowledge) proofs of correct execution would not solve the problem entirely because they only apply if the program executed entirely correctly, but this may not be the case if either the bugs that occur and cause the proof to fail are not responsible for faults (e.g., misrecording transactions) occurring, or if the program executes correctly but its logic is flawed.
Moreover proving the correct execution of a large program may simply be impractical.
Focusing on only a small critical component is not enough because if, for example, the accounting executes correctly but the display is faulty, a subpostmaster might try to fix the error manually, leading to discrepancies.

\paragraph{Uber's fraudulent activity algorithm}
In another case, the Amsterdam District Court ruled that drivers from the UK were permitted to contest the norms applied to them by Uber's system (as well as other similar companies e.g. Ola) when they were banned from the service for fraudulent activity.
Moreover, the court ordered to provide transparency about numerous aspects of its system, including the data used by Uber's algorithm to dismiss the drivers, which was not previously accessible to the drivers~\cite{ekker}.


The issue for drivers lies in the fact that they are subject to the ratings they receive from customers and Uber's system based on these customer ratings and other factors, which determine the service they receive from Uber and whether or not they are allowed to drive for Uber.
Customer ratings may be biased, however, based on attributes such as the race of the driver, which then feeds into Uber's system determining that the driver should be banned if they fall under a certain rating. 
Other surveillance systems used to assess drivers are also in place such as facial recognition checks that may fail and lead to a driver being kicked off the platform~\cite{uberface}.

Compliance based audits would not achieve much in these scenarios.
When it comes to biased customers, there is no way to assess in advance whether customers will be more or less biased, or to produce a facial recognition system that functions such that there is a negligible probability of failure across all drivers.
Inevitably, transparency will be required and must be available for drivers to allow them to contest such systems, without first having to go through lengthy, expensive processes to access the relevant information that is intentionally obfuscated.

While regulations, in this case Article 22 of the GDPR that gives an individual the right not to be subject to a decision based solely on automated processing, can enable an order to disclose aspects of the system, mechanisms to execute this are lacking, and it is not always possible for an individual to know that they are subject to such a system.
There are suggestions for ways to audit the design~\cite{raji2020closing} of AI systems as well as releasing information about the models themselves~\cite{mitchell2019model} and the datasets that they are trained on~\cite{gebru2021datasheets}. 
However, these are not designed with a threat model and, therefore, assume a fairly honest system designer and operator, whereas companies such as Uber have an incentive to obfuscate how their system functions to avoid scrutiny, and argue this is necessary for commercial confidentiality and customer privacy purposes.


\section{Practical considerations}
\subsection{Electronic evidence}
The book by Mason and Seng~\cite{mason2021electronic} discusses many issues with electronic evidence in the legal context, and makes clear that the topic touches upon many aspects of security, not only the authentication (typically handled through electronic signatures) and integrity of the evidence itself (typically handled through cryptographic hash functions), but also of the networks over which it is exchanged, and how it is stored. 
It also makes clear that when treating software as a witness, it must be taken into account that software can be written to deceive, as in the Volkswagen emissions case~\cite{volks}.

More recently, the Post Office case used as an example above has generated work discussing the presumption of reliability that evidence generated by computers often enjoy~\cite{marshall2020recommendations,ladkin2020law,ladkin2020robustness}. (Different jurisdictions adopt different standards of course.)
Related to this is also the necessity for expert witnesses to explain the evidence that is generated, so the explainability of the evidence plays an important role because the expert witness must be able to understand the evidence themselves and be able to explain it in a clear way to a judge or jury.
Explainability has been investigated for machine learning based systems, sometimes with emphasis on explaining single decisions to individual users rather than explaining a system as a whole, which may be required to establish its reliability.
The kind of explainability that is geared towards engineers~\cite{bhatt2020explainable} of the system may be more useful in this context, but may also be less accessible by design.

\subsection{Balancing transparency and privacy}
Whenever information that may be sensitive is made available, privacy and confidentiality concerns emerge.

This includes concerns for the privacy of the individuals who may be related to the information that is released.
This should be treated with care, using appropriate sanitisation mechanisms e.g., implementing data minimisation and using differentially private data release mechanisms~\cite{dwork2008differential,dwork2014algorithmic}, which are aligned with regulatory data protection requirements~\cite{cohen2020towards,nissim2017bridging}.
Because different data carries different privacy risks, and different levels of usefulness in contesting the system, this is a problem that must be addressed on a case by case basis that takes into account the trade-offs between privacy, the consent of parties that relate to the information (or other bases for releasing that information), and the information that is necessary for transparency to be useful.

Often, a dispute may rely on both system level information (e.g., error rates) and individual information (e.g., specific system events).
System level information such as univariate statistics may leak less sensitive information about individuals, while individual information is naturally more sensitive but may need only be accessible to the individual in question.

Commercial confidentiality can also be a concern.
This motivated the reliance on tools such as zero-knowledge proofs suggested by Desai and Kroll~\cite{desai2017trust} and Kroll et al.~\cite{kroll165accountable}.
There are arguments that support the idea of access to the source code of a system in the case of a dispute about the system~\cite{bellovin2021seeking}, and as we have discussed above, relying on assurances rather than transparency may not enable contestability. 
Naturally, in cases where commercial entities benefit from information asymmetry, they are unlikely to want to provide greater transparency without an incentive or obligation that would provide trade-offs in favour of transparency.
Thus, it may fall to evolving regulations and technical standards that govern the design and operation of systems to determine the right balance. 
As we have seen in the Uber example above, regulations can already force the disclosure of broad information about the system, even if they did not require that information to be public beforehand.

\subsection{A system in one place, transparency in another}
Systems are often designed and implemented in one place before being deployed internationally.
Disputes around the system, however, often take place where the harm caused by the system has occurred, which may not be where the system has been designed.
Thus, transparency around the system, if it is to be useful in a dispute, should reflect the local context of the dispute, rather than the context in which the system was designed.
The importance of the audience of transparency has been discussed by Kemper and Kolkman~\cite{kemper2019transparent} and Felzmann et al.~\cite{felzmann2019transparency}, highlighting the need for transparency solutions that reflect the population it interacts with.

\section{Related Work}
There is a vast amount of work concerned with accountability, much of which is covered in the systematization of the topic by Wieringa~\cite{wieringa2020account} that is based on Boven's framework for accountability~\cite{bovens2007analysing}.

Specifically related to this paper, there is work that focuses on security models for accountability~\cite{feigenbaum2011towards,feigenbaum2011accountability,feigenbaum2011towards}, interactions between security mechanisms that can provide assurances and the legal system have also been studied previously~\cite{kroll165accountable,desai2017trust}, as well as the production of evidence by systems~\cite{murdoch2014security}.

Our work differs from this existing body of work by considering how accountability mechanisms, in particular transparency enhancing technologies, can be used to contest norms enforced by code when designed to support existing processes such as legal disputes, rather than the predominant focus on obtaining assurances of compliance with a norm.

More recent work does address contestability, such as the work of Lyons et al.~\cite{lyons2021conceptualising} who, like us, consider the ability to contest via legal processes but focus on higher level design principles.
Our work is complementary to theirs, approaching contestability from the perspective of digisprudence and discussing specific technical security mechanisms.




\section{Conclusion}

In this paper, we have argued for the necessity of employing secure accountability mechanisms to ensure the legitimacy of computational systems whose code enforce norms.
In particular, we have argued the need for accountability mechanisms to enable the ability to contest the norms that code enforces when these may be illegitimate.
This entails a culture shift to a user centric notion aimed at giving users the agency to contest the systems they are subject to through channels such as legal processes, rather than a technical system centric notion of accountable systems that do not entail any change in systems if they are flawed.

Because the best mechanisms to contest norms are those that can effectively pressure system designers or operators to change their system even if they are reluctant to do so, such as the judicial system, we have analysed technical accountability mechanisms based on their ability to support the action of contesting computational systems via legal processes.
From this perspective, transparency enhancing technologies i.e., accountability mechanisms that include transparent logs of a system's operation, emerge as mechanisms that are more supportive of contestability than other accountability mechanisms based on providing assurances of compliance with given norms.


While work on designing technical systems has previously predominantly focused on building systems that match or comply with norms, there is scope to build upon existing tools to create better transparency enhancing technologies that fill all the requirements that must for met to effectively enable accountability and, by extension, the ability to contest and change norms.
Thus, this has implications for developers who wish to produce change in existing systems and developers of new systems that may be designed with a model of decentralised governance that affords broader scope for changing the norms enacted by the system.

While there is justified scepticism of technical solutions to governance or regulatory issues, work in the field of law and policy that is concerned with the impact of computer systems should encourage and interact with the development of technical tools that can support their goals and empower the users that their work aims to help.
Innovative new systems are not the only type of system that can cause harms, but there is necessarily a lag between technical innovation, the appearance of new systems, and of any effective governance or regulatory frameworks for these systems, which can leave users more exposed.
Until such frameworks are put in place, it is all the more important for users to be able to contest the impact new systems can have, and this can also help guide the development of these frameworks by exposing system flaws or gaps in existing regulatory and governance approaches.

\begin{acks}
Thanks to anonymous reviewers and the paper's shepherd Matt Spencer for their thoughtful comments and feedback.
In particular, I am grateful for Matt Spencer's help with the discussion of norms in Section~\ref{sec:norms}.
\end{acks}

\bibliographystyle{ACM-Reference-Format}
\bibliography{refs}

\appendix

\end{document}